\documentstyle[12pt]{article}
\def\ba{\begin{eqnarray}}
\def\ea{\end{eqnarray}}

\def\lb{\label}
\def\be{\begin{equation}}
\def\ee{\end{equation}}

\begin{document}
\title{New non compact Calabi-Yau metrics in D=6}
\author{Osvaldo P. Santillan \thanks{School of Mathematics, Trinity College, Dublin,
firenzecita@hotmail.com and osantil@math.tcd.ie}}
\date{}
\maketitle

\begin{abstract}
A method for constructing explicit Calabi-Yau metrics in six dimensions in terms of an initial hyperkahler structure is presented. The equations to solve are non linear in general, but become linear when the objects describing the metric depend on only one complex coordinate of the hyperkahler 4-dimensional space and its complex conjugated. This situation in particular gives a dual description of D6-branes wrapping a complex 1-cycle inside the hyperkahler space \cite{Fayyazuddin}. The present work generalize the construction given in that reference. But the explicit solutions we present  correspond to the non linear problem. This is a non linear equation with respect to two variables which, with the help of some specific anzatz, is reduced to a non linear equation with a single variable solvable in terms of elliptic functions.  In these terms we construct an infinite family of non compact Calabi-Yau metrics.
\end{abstract}

\tableofcontents

\section{Introduction}

    The development of the subject of Calabi-Yau (CY) manifolds is an illustrative example of the interplay between algebraic geometry and string theory.
On one hand, CY spaces are interpreted as internal spaces of string and M-theory giving supersymmetric field theories after compactification.
In fact, CY 3-folds may provide compactifications which are more realistic than the ones corresponding to other Ricci flat manifolds such as $G_2$ holonomy spaces, for
which the generation of chiral matter and non abelian gauge symmetries seems harder (but not impossible) to achieve. On the other hand, string theory compactifications
stimulated several new trends in the algebro-geometrical aspects of CY spaces, an example is the subject of mirror symmetry.

    By definition a CY manifold is a compact Kahler n-dimensional manifold with vanishing first Chern class. The Yau proof of the Calabi conjecture implies that
 these manifolds admit a Ricci flat metric and their holonomy is reduced from SO(2n) to SU(n) \cite{Yau}.  Although these compact Ricci flat metrics do exist, no explicit expression
has been found. For the non compact case, the definition usually adopted is that
a CY manifold is a Ricci flat Kahler manifold, which also implies that the holonomy is reduced to SU(n) or to a smaller subgroup. In this case, several explicit metrics are known.
One of the oldest examples are the asymptotically conical metrics presented in \cite{Candelas}.
Another interesting examples have been found in \cite{Stelle} and \cite{Rychenkova}. The last ones posses curvature singularities, but for some of them the contribution to the gravitational action
is finite and this makes plausible that they may be extended to a  six dimensional gravitational instanton. In fact, some of these solutions were
identified as the asymptotic form of the so called generalized Bando-Kobayashi-Tian-Yau (BKTY) metrics \cite{Kobayashi} and \cite{Yau2}, which are by construction Calabi-Yau. More examples were
found recently in \cite{Martelli1}-\cite{Mart} and higher dimensional ones in \cite{Vazco}-\cite{Vazquez}, some of these metrics posses conical singularities but in some cases these singularities have been resolved to give  complete metrics.

    A relatively new achievement in the subject is the characterization of the supergravity backgrounds corresponding to D6 branes wrapping a complex submanifold inside a 4-dimensional hyperkahler space. It was shown by Fayyazuddin in \cite{Fayyazuddin} that these D6 backgrounds are described in terms of a linear system of equations and the uplift to eleven dimensions results  in a purely geometrical background of the form $M_{1,4} \times Y_6$ where $Y_6$ is a Calabi-Yau space. The Ricci flat Kahler metric on $Y_6$ is therefore determined in terms of this linear system. The present work generalize this idea and is organized as follows. In  section 2 the geometry of certain Calabi-Yau spaces in six dimensions with an isometry preserving the metric and the full SU(3) structure is characterized. In addition, a class of these Calabi-Yau 3-folds constructed in terms of an initial hyperkahler structure are presented and the non linear system describing these geometries is written explicitly. It is also shown that this non linear system reduce to the linear one found in \cite{Fayyazuddin} when the objects describing the geometry depends on a single complex coordinate and its complex conjugated and also that, in accordance with
\cite{Fayyazuddin}, when the initial hyperkahler metric is the flat one on $R^4$ the resulting CY is the direct sum of $R^2$ and the multi-centered hyperkahler 4-metrics \cite{Gibhaw}. These metrics are of holonomy $SU(2)$ which is a subgroup of $SU(3)$. Section 3 and 4 improve all these situations by dealing with the full non linear problem associated to the flat hyperkahler structure. Particular solutions  are constructed and it is shown that the corresponding metrics are of holonomy exactly $SU(3)$. In addition, an infinite family of non compact Calabi-Yau metrics is presented.

\section{Calabi-Yau metrics with an isometry preserving the SU(3) structure}

\subsection{The general form of the $SU(3)$ structure}

    In this subsection a large family of Calabi-Yau (CY) manifolds with an isometry group of codimension one will be characterized. It will be assumed that the Killing vector $V$ corresponding to this isometry preserve not only the metric, but the full SU(3) structure. It will be convenient to give an operative definition of CY manifolds in six dimensions first, for more details
see for instance \cite{Joyce}-\cite{Voisin}. Roughly speaking, a Calabi-Yau manifold $M_6$ is Kahler manifold, thus complex sympletic, which in addition admits a Ricci-flat metric $g_6$. This definition means in particular that there exist an endomorphism of the tangent space $J: TM_6\to TM_6$ such that $J^2=-I_d$ and for which $g_6(X, J Y)=g_6(J X, Y)$ being $X$ and $Y$  arbitrary vector fields. It is commonly said that the metric $g_6$ is hermitian with respect to $J$ and the tensor $(g_6)_{\mu\alpha}J^{\alpha}_{\nu}$ is skew symmetric, therefore locally it defines a 2-form
\be\lb{r}
\omega_6=\frac{1}{2}\;(g_6)_{\mu\alpha}J^{\alpha}_{\nu} dx^{\mu}\wedge dx^{\nu}.
\ee
Here $x^{\mu}$ is a local choice of coordinates for $M_6$. The endomorphism $J$ it is called an almost complex structure. If the Nijenhuis tensor
$$
N(X, Y)=[X, Y]+J\;[X, J \;Y]+J\;[J\;X, Y]-[J\;X, J\;Y],
$$
vanish identically for any choice of $X$ and $Y$ then the tensor $J$ will be called a complex structure and $M_6$ a complex manifold. This is the case for a CY manifold. The Newlander-Niremberg theorem states that there is an atlas of charts for $M_6$ which are open subsets in $C^{n}$, in such a way  that the transition maps are holomorphic functions. These local charts are parameterized by complex coordinates $(z_i, \overline{z}_i)$ with $i=1,2,3$ for which the complex structure looks like
\be\lb{lula}
J_{i}^{j}=-J_{\overline{i}}^{\overline{j}}=i\delta_i^j,\qquad J_{i}^{\overline{j}}= J^{i}_{\overline{j}}=0,
\ee
and for which the metric and the 2-form (\ref{r}) are expressed as follows
\be\lb{expro}
g_6=(g_6)_{i\overline{j}} \;dz_{i} \otimes d\overline{z}_{j},
\ee
\be\lb{expro2}
\omega_6=\frac{i}{2}\;(g_6)_{i\overline{j}} \;dz_{i}\wedge d\overline{z}_{j}.
\ee
The form (\ref{expro2}) is called of type $(1,1)$ with respect to $J$, while a generic 2-form containing only terms of the form $(dz_i\wedge dz_j)$ or $(d\overline{z}_i\wedge d\overline{z}_j)$ will be called of type $(2, 0)$ or $(0, 2)$, respectively.  In addition a Calabi-Yau manifold is sympletic with respect to $\omega_6$, in other words $d\omega_6=0$. A complex manifold which is sympletic with respect to (\ref{r})  is known as Kahler manifold, thus CY spaces  are all Kahler. The Kahler condition itself implies that the holomy is reduced from $SO(6)$ to $U(3)$. Furthermore, the fact that $g_6$ is Ricci-flat is equivalent to the existence of a 3-form
\be\lb{s}
\Psi=\psi_{+}+i \;\psi_{-},
\ee
of type $(3,0)$ with respect to $J$, satisfying the following compatibility conditions \cite{Chiozzi}
\be\lb{condi}
\omega_6\wedge \psi_{\pm}=0,\qquad \psi_{+}\wedge \psi_{-}=\frac{2}{3}\omega_6\wedge \omega_6\wedge \omega_6\simeq dV(g_6),
\ee
$$
\psi_{+}\wedge \psi_{+}=\psi_{-}\wedge \psi_{-}=0.
$$
in such a way that
\be\lb{cob}
d\psi_{+}=d\psi_{-}=0.
\ee
In the formula (\ref{condi}) $dV(g_6)$ denote the volume form of $g_6$. In the situations described in (\ref{cob})  the holonomy is further reduced from $U(3)$ to $SU(3)$, thus CY manifolds are of $SU(3)$ holonomy. The converse of these statements are also true, that is, for any Ricci flat Kahler metric in D=6 there will exist an SU(3) structure $(\omega_6, \Psi)$ satisfying (\ref{condi}) and also
\be\lb{co}
d\omega_6=d\psi_{+}=d\psi_{-}=0.
\ee
The knowledge SU(3) structure determine univocally metric $g_6$. In fact, the task to find complex coordinates for a given CY manifold may be not simple, but there always exists a tetrad basis $e^a$
with $a=1,..,6$ for which the SU(3) structure is expressed as
$$
\omega_6=e^1\wedge e^2+e^3\wedge e^4+e^5\wedge e^6,
$$
\be\lb{bubu}
\psi_+=(e^1\wedge e^4+e^2\wedge e^3)\wedge e^5+(e^1\wedge e^3+e^4\wedge e^2)\wedge e^6,
\ee
$$
\psi_-=-(e^1\wedge e^3+e^4\wedge e^2)\wedge e^5+(e^1\wedge e^4+e^2\wedge e^3)\wedge e^6,
$$
and for which the metric is $g_6=e^a\otimes e^a$, where the Einstein summation is understood.

      The description given above just collect general facts about CY manifolds. In the following we will assume that our CY manifold $M_6$ is equipped with a metric $g_6$ in such a way that there is a Killing vector $V$ preserving $g_6$ and the whole SU(3) structure ($\omega_6$, $\psi_+$, $\psi_-$). \footnote{In fact, a vector V preserving the whole $SU(3)$ structure automatically preserve $g_6$.} For this situation there exists a local coordinate system $(\alpha, x^i)$ with $i=1,..,5$ for which $V=\partial_{\alpha}$ and for which the metric tensor $g_6$  take the following form
\be\lb{CY}
g_6=\frac{(d\alpha+A)^2}{H^2}+H g_5.
\ee
where the function $H$, the one form $A$ and the metric tensor $g_5$ are independent on the coordinate $\alpha$. Thus these objects live in a 5-dimensional space
which we denote $M_5$.  The metric $g_5$ appearing in (\ref{CY}) can be expressed as $g_5=e^a\otimes e^a$ with $a=1,..,5$ for some basis of $\alpha$-independent 1-forms $e^a$. Then,
if $V$ also preserve the $SU(3)$ structure (as we are assuming) one has the decomposition
\be\lb{su3}
\omega_6=\omega_1+\frac{1}{\sqrt{H}}e^5\wedge (d\alpha+A),
\ee
\be\lb{su32}
\psi_{+}=H^{3/2}\omega_3\wedge e^5+\omega_2\wedge (d\alpha+A),
\ee
\be\lb{su33}
\psi_{-}=-H^{3/2}\omega_2\wedge e^5+\omega_3\wedge (d\alpha+A).
\ee
Here the 1-form $e^5$ and the two-forms $\omega_2$ and $\omega_3$ are by definition
\be\lb{analog1}
\frac{e^5}{\sqrt{H}}=i_{\partial_{\alpha}}\omega_{6},
\ee
\be\lb{mariga}
\omega_2=i_{\partial_{\alpha}}\psi_{+},\qquad \omega_3=i_{\partial_{\alpha}}\psi_{-}.
\ee
$i_{V}$ denoting the contraction with the vector field $V$. In fact the triplet of forms $\omega_i$ may be represented as $\omega_i=e^4\wedge e^i+\epsilon_{ijk}e^j\wedge e^k$ with $i=1,2,3$ for some convenient choice of $e^a$. By assumption $V$ preserve (\ref{su3})-(\ref{su33}) and therefore $e^5$ and $\omega_i$ do not depend on the coordinate $\alpha$, i.e, they are also defined in $M_5$.

    The task now is to derive the consequences of the CY relations (\ref{condi}) and (\ref{co}) for the generic anzatz (\ref{CY})-(\ref{su33}).  From (\ref{condi}) it follows immediately that
\be\lb{conse}
\omega_1\wedge\omega_2=\omega_1\wedge \omega_3=\omega_2\wedge\omega_3=0,
\ee
\be\lb{conse2}
\omega_1\wedge\omega_1=H^2\omega_2\wedge\omega_2=H^2\omega_3\wedge\omega_3.
\ee
Moreover (\ref{analog1}) together with an elementary formula in differential geometry imply that
\be\lb{analog2}
d_5(\frac{e^{5}}{\sqrt{H}})=\pounds_{\partial_{\alpha}}\omega_6-i_{\partial_{\alpha}}\;d\omega_6.
\ee
Here $d_5=\partial_i\; dx^i$ and $\pounds_{\partial_{\alpha}}$ is the Lie derivate along the vector $\partial_{\alpha}$. But the vector $\partial_{\alpha}$, by assumption, preserves $\omega_6$ and $\omega_6$ is closed, thus the right hand side of the last expression vanish and
\be\lb{s2}
d_5(\frac{e^{5}}{\sqrt{H}})=0.
\ee
The last relation can be integrated, at least locally, to obtain that
\be\lb{retu}
e^5=\sqrt{H}\;dy,
\ee
$y$ being some function of the coordinates $x^i$ parameterizing $M_5$, which is the momentum map of the isometry.  At least locally, one can take the function $y$ defined in (\ref{retu}) as one of the coordinates, which leads to the decomposition $M_5=M_4\times R_{y}$ and $d_5=d_4+\partial_y \;dy$. The metric (\ref{CY}) in this coordinates becomes
\be\lb{CY2}
g_6=\frac{(d\alpha+A)^2}{H^2}+H^2 dy^2+g(y),
\ee
where $g(y)$ will be determined below under certain additional assumptions. An analogous calculation taking into account (\ref{mariga}) shows that
\be\lb{cluso}
d_5\omega_2=d_5\omega_3=0.
\ee
In the remaining part of the paper it will be assumed that the forms $\omega_i$ are defined on $M_4$ and that they depend on $y$ as a parameter. We are not sure if this is the most general case, but is the one that we were able to deal with. Formally, this means that
\be\lb{susu}
\omega_i(X, \partial_y)=0.
\ee
 But the closure of $\omega_2$ and $\omega_3$ (\ref{cluso}) together with the decomposition $d_5=d_4+\partial_y \;dy$ and (\ref{susu}) imply that $\omega_2$ and $\omega_3$ are $y$ independent. Thus only $\omega_1$ is allowed to depend on $y$. In addition $g(y)$ can be interpreted as a four dimensional metric depending on $y$ as a parameter. Taking into account this considerations, the $SU(3)$ structure (\ref{su3}) and (\ref{su33}) takes the following form
$$
\omega_6=\omega_1(y)+dy\wedge (d\alpha+A),
$$
\be\lb{nova}
\psi_{+}=H^{2}\omega_3\wedge dy+\omega_2\wedge (d\alpha+A),
\ee
$$
 \psi_{-}=-H^{2}\omega_2\wedge dy+\omega_3\wedge (d\alpha+A).
$$
The next task will be to find specific examples of this type of Calabi-Yau structures.

\subsection{Solutions related to hyperkahler 4-dimensional spaces}

  The method we will employ in order to find particular Calabi-Yau structures of the form (\ref{nova}) is to start with an hyperkahler structure $\widetilde{\omega}_i$ defined over a 4-manifold $M_4$. As is well known, a 4-dimensional hyperkahler manifold is a Kahler one which admits a Ricci flat metric $g_4$. This automatically imply  that the holonomy is in $SU(2)$. In fact, these manifolds can be considered as 4-dimensional CY spaces; they are complex with respect to some endomorphism $J_1$ of the tangent space and that the corresponding form $\widetilde{\omega}_1$ is of type (1,1) with respect to $J_1$ and closed, i.e, $d\widetilde{\omega}_1=0$.  The reduction of the holonomy to $SU(2)$ imply the existence of a complex 2-form $\widetilde{\omega}_2+i\;\widetilde{\omega}_3$ is of type (2,0) with respect to $J_1$ and so
\be\lb{binban}
\widetilde{\omega}_1\wedge \widetilde{\omega}_2=\widetilde{\omega}_1\wedge \widetilde{\omega}_3=0,
\ee
and which, in addition, satisfy
\be\lb{enya}
\widetilde{\omega}_1\wedge \widetilde{\omega}_1=\widetilde{\omega}_2\wedge \widetilde{\omega}_2 =\widetilde{\omega}_3\wedge \widetilde{\omega}_3\simeq dV(g_4),
\ee
\be\lb{binban2}
d_4\widetilde{\omega}_2=d_4\widetilde{\omega}_3=0.
\ee
The form $\widetilde{\omega}_2+i\;\widetilde{\omega}_3$ is in fact the analog of the 3-form $\Psi$ defined in (\ref{s}) for the 4-dimensional case. The conditions stated above imply that all the endomorphisms $J_i$  defined through the usual relation
\be\lb{usu}
g_4(X, J_i\;Y)=\widetilde{\omega}_i(X, Y),
\ee
are all complex structures and that $M_4$ is Kahler with respect to any $\widetilde{\omega}_i$. In fact any endomorphism of the form
\be\lb{bundo}
J=a_1 \;J_1+a_2\; J_2+a_3\;J_3, \qquad a^2_1+a^2_2+a_3^2=1,
\ee
is a complex structure as well, so there is an $S^2$-bundle of complex structures for any hyperkahler manifold. The triplet of Kahler forms $\widetilde{\omega}_i$ is enough to determine the hyperkahler metric $g_4$ univocally, as in the six dimensional case.

     Given an hyperkahler structure $\widetilde{\omega}_i$, we deform $\widetilde{\omega}_1$ to a $y$-dependent two form
\be\lb{der}
\omega_1(y)=\widetilde{\omega}_1-d_4 d_4^c G,
\ee
while keeping $\widetilde{\omega}_2$ and $\widetilde{\omega}_3$ intact. Here we have introduced the operator $d^c=J_1\;d$. In the expression (\ref{der})  $G$ denotes an unknown function
which varies on $M_4$ and depends also on $y$.  In terms of  this anzatz the SU(3) structure (\ref{nova})  take the following form
$$
\omega_6=\widetilde{\omega}_1-d_4 d^c_4 G+dy\wedge (d\alpha+A),
$$
\be\lb{nova2}
\psi_{+}=H^{2}\widetilde{\omega}_3\wedge dy+\widetilde{\omega}_2\wedge (d\alpha+A),
\ee
$$
\psi_{-}=-H^{2}\widetilde{\omega}_2\wedge dy+\widetilde{\omega}_3\wedge (d\alpha+A).
$$
The reason for the choice (\ref{der}) for $\omega_1(y)$ is  simple to explain.  The form $\omega_1$ in (\ref{der}) is of type (1,1) with respect to the complex coordinates which diagonalize $J_1$, and the term $d_4 d_4^c G$ is also of this type. Thus the deformed form $\omega_1(y)$ is of type (1,1) as well. As the form $\widetilde{\omega}_2+i\;\widetilde{\omega}_3$ is kept intact and is of type $(2,0)$ the fundamental condition (\ref{conse}) is identically satisfied. Additionally (\ref{der}) is  closed with respect to $d_4$, and this will simplify considerably the analysis given below.

      Given the deformed structure (\ref{der}), the compatibility condition (\ref{conse2}) imply that
\be\lb{dofini}
(\widetilde{\omega}_1-d_4 d^c_4 G)\wedge(\widetilde{\omega}_1-d_4 d^c_4 G)=H^2 \widetilde{\omega}_2\wedge \widetilde{\omega}_2.
\ee
But the wedge products appearing in the last equality are all proportional to the volume form $dV(g_4)$ of the initial hyperkahler metric $g_4$,  therefore the relation
\be\lb{defini}
(\widetilde{\omega}_1-d_4 d^c_4 G)\wedge(\widetilde{\omega}_1-d_4 d^c_4 G)=M(G)\;\widetilde{\omega}_1\wedge \widetilde{\omega}_1,
\ee
defines a non linear expression $M(G)$ involving $G$.  In the following we will loosely call $M(G)$ a non linear operator,
although the correct spelling is that $M(G)$ is the action of a non linear operator over $G$. We will use this terminology in order to simplify the vocabulary, when there is no place for confusion.
Taking into account the two last formulas (\ref{dofini}) and (\ref{defini}) together with (\ref{enya}) it follows that
\be\lb{munja}
M(G)=H^2.
\ee
The CY condition (\ref{co}) applied to (\ref{nova2}) impose further constraints. The closure of $\omega_6$
$$
d\omega_6=d\omega_1-dy\wedge (d_4 d^c_4 G_y+ d_4A)=0,
$$
together with the Kahler condition $d\omega_1=0$ imply that
\be\lb{sancher}
A=-d^c_4 G_y,
 \ee
 up to a gauge transformation that can be absorbed by a redefinition of the coordinate $\alpha$ in (\ref{nova2}). Furthermore, the closure of $\psi_{\pm}$ gives the following equations
\be\lb{sancher2}
\widetilde{\omega}_2\wedge d_4A=\widetilde{\omega}_3\wedge d_4A=0,\ee
\be\lb{sancher9}
\widetilde{\omega}_3\wedge d_4H^2\wedge dy+\widetilde{\omega}_2\wedge dy \wedge A_y=0.
\ee
But the  condition (\ref{sancher2})  is identically satisfied by use of (\ref{sancher}) and therefore redundant, as
$$
\widetilde{\omega}_2\wedge d_4A=-\widetilde{\omega}_2\wedge d_4 d^c_4 G_y=0,
$$
because $\omega_2$ contain only terms of type $(2,0)$ and $(0,2)$ with respect to $J_1$ while $d_4 d^c_4 G_y$ is purely of type (1,1). Instead (\ref{sancher9})  give new constraints. The two forms $\widetilde{\omega}_2$ and $\widetilde{\omega}_3$ are related by
\be\lb{almcomp}
\widetilde{\omega}_2(X, Y)=\widetilde{\omega}_3(J_1 X, Y),
\ee
and it is a general fact that an equation of the form
$$
\widetilde{\omega}_2\wedge \alpha+\beta \wedge \widetilde{\omega}_3=0,
$$
is solved when
$$
\alpha=J_1 \beta.
$$
Taking into account this and the second (\ref{sancher2}) we conclude that
\be\lb{mer}
d_4^c H^2=-\partial_y A.
\ee
From (\ref{mer}) combined with (\ref{sancher}) it follows that
\be\lb{dure}
G_{yy}=H^2.
\ee
From the last equation together with (\ref{munja}) it is obtained a non linear differential equation determined the function $G$, namely
\be\lb{munja2}
G_{yy}=M(G).
\ee
This is the fundamental equation we will solve along the text. Finally, we have checked that the closure of $\psi_{-}$ gives no new constraints, but we omit this analysis.
By introducing (\ref{sancher9}) and  (\ref{dure}) into (\ref{nova2}) it follows that the SU(3) structure is
$$
\omega_6=\widetilde{\omega}_1-d_4 d^c_4 G+dy\wedge (d\alpha-d^c_4 G_y ),
$$
\be\lb{nova3}
\psi_{+}=G_{yy}\;\widetilde{\omega}_3\wedge dy+\widetilde{\omega}_2\wedge (d\alpha-d^c_4 G_y ),\ee
$$
\psi_{-}=-G_{yy}\;\widetilde{\omega}_2\wedge dy+\widetilde{\omega}_3\wedge (d\alpha-d^c_4 G_y ),
$$
which can be checked to be closed. The generic form of the 6-dimensional Calabi-Yau metric corresponding to this structure is given by
\be\lb{gonoro}
g_6=g_{4}(y)+G_{yy}\;dy^2+\frac{(d\alpha-d^c_4 G_y )^2}{G_{yy}},
\ee
where $g_{4}(y)$  is  the Kahler 4-dimensional metric corresponding to the deformed Kahler structure $\omega_1(y)=\widetilde{\omega}_1-d_4 d^c_4 G$. Note that $g_4(y)$ depends on $y$ as a parameter, in other words, it is a four dimensional Kahler metric which varies as $y$ take different values.

\subsection{The Fayyazuddin linearization}

   It is important to remark that the family of SU(3) structures (\ref{nova3}) and (\ref{gonoro}) found above are completely determined in terms of a single function $G$ and derivates, therefore
the task to find our CY metrics have been reduced to solve a single equation (\ref{munja2}) defining $G$. This is a non linear equation and the general solution
is not known, but it can be solved in some specific examples. We will focus now our attention in particular solutions in this equation. The source of the non linearity of the operator $M(G)$ defined in (\ref{defini}) and (\ref{munja2}) is the quadratic term
\be\lb{bedford}
Q(G)=d_4 d_4^c G\wedge d_4 d_4^c G,
\ee
therefore the operator $M(G)$ will reduce to a linear one if $Q(G)$ vanish.  This will be the case when the function $G$ is of the form $G=G(w, \overline{w})$ where $w=f(z_1, z_2)$ is an holomorphic
function of the coordinates $(z_1, z_2)$ which diagonalize the complex structure $J_1$ \cite{Bedford}. This affirmation may be justified as follows. From the formula
\be\lb{df}
d d^cG=2 \;i \;G_{i\overline{j}}\;dz_i\wedge d\overline{z}_j,
\ee
 it follows that (\ref{bedford}) is
\be\lb{game}
Q(G)=-8\;(G_{1\overline{1}}G_{2\overline{2}}-G_{1\overline{2}}G_{2\overline{1}})\; dz_1\wedge d\overline{z}_1\wedge dz_2\wedge d\overline{z}_2.
\ee
But the functional dependence  $G=G(w, \overline{w})$ imply that
$$
G_{i\overline{j}}= w_i\; \overline{w}_{\overline{j}}\;G_{w\overline{w}},
$$
and by inserting this into (\ref{game}) gives $Q(G)=0$. This result may be paraphrased as follows. If the function $G$ depends only on two complex coordinates $(w, \overline{w})$ then the quantity $d_4 d_4^c G$ is essentially a 2-form in two dimensions, therefore the wedge product (\ref{bedford}) vanish identically.

      The situation described above is essentially the one considered by Fayyazuddin in the reference \cite{Fayyazuddin} and, if suitable boundary conditions are imposed,  the resulting metrics give a dual description of D6 branes wrapping a complex submanifold in a hyperkahler manifold.  A simple example is obtained when the initial hyperkahler structure is the flat metric on $R^4$
\be\lb{flot}
g_4=dz_1\otimes d\overline{z}_1+dz_2\otimes d\overline{z}_2,
\ee
with the Kahler triplet which is expressed in complex form as
\be\lb{1}
\widetilde{\omega}_1=\frac{i}{2}(dz_1\wedge d\overline{z}_1+dz_2\wedge d\overline{z}_2),
\ee
\be\lb{2}
\widetilde{\omega}_2+i\widetilde{\omega}_3=dz_1\wedge dz_2.
\ee
The volume 4-form for (\ref{flot}) in complex coordinates is simply
\be\lb{int}
\widetilde{\omega}_1\wedge \widetilde{\omega}_1=-\frac{1}{2}\;dz_1\wedge d\overline{z}_1\wedge dz_2\wedge d\overline{z}_2.
\ee
In these terms the relation (\ref{defini}) for the flat metric in $R^4$ takes the following form
\be\lb{ma}
(\widetilde{\omega}_1-d d^cG) \wedge (\widetilde{\omega}_1-d d^cG)=-\frac{1}{2}\;M(G)\;dz_1\wedge d\overline{z}_1\wedge dz_2\wedge d\overline{z}_2.
\ee
In addition the formula (\ref{df}) obtained previously gives the linear term
$$
2\;d d^cG \wedge \widetilde{\omega}_1=4\;(G_{1\overline{1}}+G_{2\overline{2}}) \;\widetilde{\omega}_1\wedge \widetilde{\omega}_1,
$$
and also the following quadratic one
$$
Q(G)=dd^cG \wedge d d^cG=16\;(G_{1\overline{1}}G_{2\overline{2}}-G_{1\overline{2}}G_{2\overline{1}})\;\widetilde{\omega}_1\wedge\widetilde{\omega}_1.
$$
By use of the last two identities it is obtained from (\ref{ma}) the following expression for $M(G)$
\be\lb{ss}
M(G)=1-4\;(G_{1\overline{1}}+G_{2\overline{2}}) +16\;(G_{1\overline{1}}G_{2\overline{2}}-G_{1\overline{2}}G_{2\overline{1}}).
\ee
The fundamental equation  defining the Calabi-Yau geometry is then obtained from (\ref{munja2}) and (\ref{ss}), the result is
\be\lb{suv}
G_{yy}=1-4\;(G_{1\overline{1}}+G_{2\overline{2}}) +16\;(G_{1\overline{1}}G_{2\overline{2}}-G_{1\overline{2}}G_{2\overline{1}}).
\ee
The next problem is to find particular solutions of (\ref{suv}). As was discussed above if one assume that the solution depends on one complex coordinate $w=f(z_i)$, then the non linear term in (\ref{suv}) vanish identically and the resulting equation is
\be\lb{suv2}
G_{yy}=1-4\;(G_{1\overline{1}}+G_{2\overline{2}}).
\ee
This is a Laplace equation in five dimensions parameterized by $(y, z_1, z_2, \overline{z}_1, \overline{z}_2)$. We may derivate (\ref{suv2}) twice with respect to $y$ which gives a Laplace equation for the quantity $V=G_{yy}$ defining the radius of the "circle" in (\ref{gonoro}). This equation is
\be\lb{suvenir}
V_{yy}+4\;(V_{1\overline{1}}+V_{2\overline{2}})=0.
\ee
In the following we choose $w=z_1$.  A solution with good behavior at infinite is obtained when we put  constant density charge at the planes $z_1=c_i$, in other words the right hand side of (\ref{suvenir}) is zero up to a delta term of the form $\delta \rho=\sum_{i=1}^N q\;\delta(z_1-c_i)\;\delta(\overline{z}_1-\overline{c}_i)\;\delta(y-a_i)$. The resulting electrostatic potential is
\be\lb{pet}
V=1 +\sum_{i=1}^N \frac{q}{\sqrt{|z^1 -c_i|^2 + (y-a_i)^2}},
\ee
and it is commonly said that the D6 branes are wrapped on the $z_2$ plane. We can also find the Kahler potential $G$ explicitly through the relation $V=G_{yy}$ by a double integration
\be\lb{pit2}
G= 2\;q\;\sum_{i=1}^N \bigg[\;(y-a_i)\;\ln\{(y-a_i)+\sqrt{(y-a_i)^2+|z^1-c_i|^2})\}-\sqrt{(y-a_i)^2+|z^1-c_i|^2}\;\bigg].
\ee
The metric $g_4(y)$ corresponding to (\ref{pit2}) is given by
\be\lb{ververfer}
g_{1\bar 1} =(1 +G_{1\overline{1}})\; dz_1\otimes d\overline{z}_1+(1 +G_{2\overline{2}})\; dz_2\otimes d\overline{z}_2
= V dz_1\otimes d\overline{z}_1+dz_2\otimes d\overline{z}_2.
\ee
The full Calabi-Yau metric (\ref{gonoro}) take the following form
\be\lb{improve}
g_6= dz_2\otimes d\overline{z}_2 +V\;(dy^2 +  dz_1\otimes d\overline{z}_1) + \frac{1}{V}\;(d\psi + A_1 dz^1 + A_{\bar 1}dz^{\bar 1})^2,
\ee
where $A$ is may be calculated  by use of  (\ref{sancher}) and the result satisfy
\be\lb{gier}
\nabla \times A=\pm\; \nabla V.
\ee
Therefore (\ref{improve}) is the direct sum of a flat metric in $R^2\simeq C$ and a general Gibbons-Hawking metric in dimension four \cite{Gibhaw}. As the Gibbons-Hawking metrics are hyperkahler the whole metric (\ref{improve}) is of holonomy SU(2), which is a subgroup of SU(3). Our aim in the following is to improve this situation and find Calabi-Yau metrics of holonomy \emph{exactly} SU(3).

\section{Non trivial Calabi-Yau extensions of the flat metric}

   As we have shown above, the Fayyazuddin linearization is easy to perform for the flat hyperkahler metric on $R^4$. For other hyperkahler structures the task is much harder, and we did not manage to find an explicit solution in those cases. This does not means they do not exist, but that is difficult to find them, at least for us. The problem is that when the initial hyperkahler structure has non trivial curvature the laplace type equation (\ref{suv2}) is generalized to some equation which is schematically
\be\lb{edu}
g^{i\overline{j}}\;G_{i\overline{j}}+G_{yy}=1,
\ee
up to constants which are irrelevant for our discussion.  Here $g^{i\overline{j}}$ is the inverse metric of the initial hyperkahler structure. After writing this equation explicitly one assume a functional dependence of the form $G(w,\overline{w})$ with some guess $w=f(z_i)$ and then try to solve it. The existence of solutions of (\ref{edu}) of the form $G(w,\overline{w})$ is insured only if
\be\lb{edu2}
g^{i\overline{j}}\;w_{i}\overline{w}_{\overline{j}}=L(w,\overline{w}),
\ee
for our guess $w$, $L(w,\overline{w})$ being a function which depends only on $(w,\overline{w})$. But the inverse metric $g^{i\overline{j}}$ of the initial hyperkahler structure depends in general on two complex variables, say $(w, z_2)$ and their complex conjugates. Therefore it may be very hard to find a clever guess in order to assure that the left hand side of (\ref{edu2}) is $z_2$ independent. That is the main technical problem to find Calabi-Yau metrics by means of the Fayyazuddin linearization.

   The difficulty described above should not be interpreted as a no-go theorem for the Fayyazuddin linearization. In fact we suspect that non trivial solutions may be found for some hyperkahler manifolds.  In fact, it has been pointed out in \cite{Gomis}-\cite{Edelstein2} that the supergravity solution describing a D6-branes wrapping the $S^2$ in Eguchi-hanson gravitational instanton is described by the resolved conifold metric \cite{Candelas}. Arguably the Fayyazuddin linearization may work for this instanton, we hope to elaborate this point in a future, at the moment we have no answer.

   The last alternative is to assume that $G$ do not satisfy the constraint (\ref{edu2}). In this case the equation to solve is the non linear one (\ref{munja2}) and this may be even a harder task.
The facts explained above seem to be discouraging when one attempts to find a 6-dimensional metric with holonomy exactly SU(3) by our methods. But in the remaining part of this work it will be shown that some solutions can be found explicitly. The idea behind is to tackle the full non linear equation (\ref{munja2}) for the flat hyperkahler structure on $R^4$. In fact, we worked out explicitly this equation in (\ref{suv}). Particular solutions will be found below and it will be shown that some of them correspond to non compact Ricci flat Kahler metrics with holonomy exactly SU(3).

\subsection{The "radial" anzatz}

   A simple guess for solving the fundamental non linear equation (\ref{suv}) or equivalently
 \be\lb{red}
G_{yy}-1+4\;(G_{1\overline{1}}+G_{2\overline{2}}) -16\;(G_{1\overline{1}}G_{2\overline{2}}-G_{1\overline{2}}G_{2\overline{1}})=0.
\ee
is to take $G$ is a function of $y$ and the "radius" $u=r^2=z_1\overline{z}_1+z_2\overline{z}_2$. In this situation the Fayyazuddin linearization does not work, as $G$ is not of the form $G(w, \overline{w})$ with $w$ an holomorphic function of $z_i$. For this anzatz for $G$ the following identities are easy to prove
\be\lb{azuma}
G_{i}=\overline{z}_i \;G_{u},\qquad G_{\overline{i}}=z_i\; G_{u},\qquad
G_{i\overline{j}}=\delta_{ij} \;G_u+z_i\;\overline{z}_i\; G_{uu},
\ee
from where it is obtained that
$$
G_{1\overline{1}}+G_{2\overline{2}}=2\;G_u+u\;G_{uu},
$$
and also that
$$
G_{1\overline{1}}G_{2\overline{2}}-G_{1\overline{2}}G_{2\overline{1}}=(G_u+z_1\;\overline{z}_1\; G_{uu}) \;(G_u+z_2\;\overline{z}_2\; G_{uu})-z_1\;\overline{z}_1\;z_2\;\overline{z}_2\; G_{uu}
$$
$$
=(G_u)^2+u\;G_{uu}\;G_{u}.
$$
With the help of the last equalities the equation (\ref{red}) becomes
\be\lb{edo}
G_{yy}-1-16 \;(G_u)^2-16\;u\;G_{u}\;G_{uu}+8 \;G_{u}+4\;u\;G_{uu}=0,
\ee
or, by rearranging terms
\be\lb{ed}
G_{yy}-1-8 \;(2\;G_u-1)\;(G_u+u\;G_{uu})-4\;u\;G_{uu}=0.
\ee
A simple solution of (\ref{ed}) is found by further postulating the following quadratic expression for $G$
\be\lb{postu}
G=\frac{a}{2}\;u^2+b\;u+c,
\ee
in which the coefficients $a$, $b$ and $c$ are assumed to be functions of $y$. By introducing this anzatz into (\ref{ed}) it is obtained that
\be\lb{ear}
\frac{a_{yy}}{2}\;u^2+b_{yy}\;u+c_{yy}-8\;(2\;a\;u+2\;b-1)\;(2\;a\;u+b)-4\;a\;u-1=0.
\ee
This is a polynomial expression in $u$ and it will be identically zero if its coefficients vanish identically. This lead to three equations which can be written in the following form
$$
a_{yy}-64\;a^2=0,\qquad
(b-\frac{1}{4})_{yy}-48(b-\frac{1}{4})\;a=0,\qquad c_{yy}-16\;(b-\frac{1}{2})\;b-1=0.
$$
By introducing the quantity $\xi=b-1/4$ the last system can be expressed as
\be\lb{van1}
a_{yy}-64\;a^2=0,\qquad
\xi_{yy}-48\;\xi\;a=0,\qquad c_{yy}-16\;\xi^2=0.
\ee
The solutions of the system (\ref{van1}) will determine a family of  Calabi-Yau geometries.

\subsection{The general solution of the defining equations}

  The interesting fact about the procedure implemented above is that we have reduced a non linear equation in two variables ($y$, $u$) namely, (\ref{red}), to a non linear system in one variable $y$  with three unknown functions ($a(y)$, $b(y)$, $c(y)$) namely, (\ref{van1}). The unique non linear equation in (\ref{van1}) is the first one defining $a(y)$ and this turns out
to be very easy to solve in terms of elliptic functions. Once this equation is solved, the remaining ones are straightforward. The first equation (\ref{van1}) can be easily integrated to give
\be\lb{uve}
a_{y}=\pm\sqrt{m+\frac{128\;a^3}{3}},
\ee
with $m$ a parameter with arbitrary values. We will consider two cases separately.  For $m>0$ the equation (\ref{uve}) implies that
$$
y(a)=\pm\frac{1}{\sqrt{m}}\int_{a}^{\infty} \frac{dv}{\sqrt{1+\frac{128\;v^3}{3\;m}}}+y_0.
$$
With the help of the formula
\be\lb{verve}
\int_{-\infty}^c\frac{dx}{\sqrt{1-x^3}}=\frac{1}{\sqrt{3}} F(\phi, k),
\ee
with
$$
\phi=\cos^{-1}\bigg(\frac{c-1+\sqrt{3}}{c-1-\sqrt{3}}\bigg),\qquad k=\sin\frac{5\;\pi}{12},
$$
found in section 1.2.70 of \cite{Prudnikov} it follows after a short calculation that
\be\lb{verver2}
y(a)=\pm \frac{1}{\sqrt{3\;m}}\;\bigg(\frac{3\;m}{128}\bigg)^{\frac{1}{3}} F(\phi(a), k)+y_0,
\ee
$$
\phi(a)=\cos^{-1}\bigg(\frac{(128)^{\frac{1}{3}}\;a+(3\;m)^{\frac{1}{3}}-(3\;m)^{\frac{1}{3}}\;\sqrt{3}}{(128)^{\frac{1}{3}}\;a+(3\;m)^{\frac{1}{3}}+(3\;m)^{\frac{1}{3}}\;\sqrt{3}}\bigg),\qquad k=\sin\frac{5\;\pi}{12}.
$$
Here $F(\phi, k)$ is an elliptic function of the first kind. The formula (\ref{verver2}) gives $y$ as a function of $a$. Note that the range of $y$ is finite, as the function $F$ take values in a finite range. The range of $a$ is bounded from below but not from above, that is
\be\lb{belo}
-1<\frac{128\;a^3}{3\;m},
\ee
and it follows that $a$ can take positive and negative values as well. When $m<0$ the relation $y(a)$ is given by
$$
y(a)=\pm\frac{1}{\sqrt{s}}\int_{a}^{\infty} \frac{dv}{\sqrt{\frac{128\;v^3}{3\;s}-1}}+y_0,
$$
where $s$ is defined as $s=-m$. As for the case corresponding to $m>0$, the value of $a$ is bounded from below and not from above
\be\lb{belo2}
1<\frac{128\;a^3}{3\;s},
\ee
but the difference is that $a$ can take only positive values.  The solution in this case is given by
\be\lb{verver3}
y(a)=\pm \frac{1}{\sqrt{3\;s}}\;\bigg(\frac{3\;s}{128}\bigg)^{\frac{1}{3}} F(\psi(a), k)+y_0,
\ee
$$
\psi(a)=\cos^{-1}\bigg(\frac{(128)^{\frac{1}{3}}\;a+(3\;s)^{\frac{1}{3}}+(3\;s)^{\frac{1}{3}}\;\sqrt{3}}{(128)^{\frac{1}{3}}\;a+(3\;s)^{\frac{1}{3}}-(3\;s)^{\frac{1}{3}}\;\sqrt{3}}\bigg),\qquad k=\sin\frac{\pi}{12}.
$$
The formulas (\ref{verver2}) and (\ref{verver3}) are the full solution of the first equation (\ref{van1}).

      The second  equation (\ref{van1}) gives $\xi$ as function of $y$, but it seems to be a bit more problematic, as to solve it requires
to invert (\ref{verver2}) or (\ref{verver3}) and to express $a$ as a function of $y$.  Instead of doing it is more direct to calculate $\xi$ as a function of $a$, which implicitly give it as function of $y$ by (\ref{verver2}) or (\ref{verver3}). From an elementary chain rule using (\ref{uve}) it is easy to show that, for an arbitrary function $f(y)$ of the variable $y$, the following identities are true
$$
f_y=a_y\;f_a=\pm\sqrt{m+\frac{128}{3}\;a^3}\;f_{a},\qquad f_{yy}=(m+\frac{128}{3}\;a^3)\;f_{aa}+64\;a^2\;f_{a}.
$$
In these terms the second (\ref{van1}) is rewritten as
\be\lb{ouate2}
(m+\frac{128}{3}\;a^3)\;\xi_{aa}+64\;a^2\;\xi_{a}-48\;a\;\xi=0.
\ee
By further introducing the natural variable
\be\lb{grt}
z=-\frac{128\;a^3}{3\;m},
\ee
it follows after some calculation that the second equation (\ref{van1}) is
\be\lb{hoerv}
z\;(1-z)\; \xi_{zz}+(\frac{2}{3}-\frac{7}{6}\;z)\;\xi_z+\frac{1}{8}\; \xi=0.
\ee
This is an hypergeometric equation, namely, an equation of the form
\be\lb{ouate3}
z\;(1-z)\;\xi_{zz}+[n-(p+q+1)\;z]\;\xi_{z}-p\;q\;\xi=0,
\ee
with the following values for the constants $n$, $p$ and $q$
$$
p=\frac{1}{12}-\frac{\sqrt{19}}{12},\qquad q=\frac{1}{12}+\frac{\sqrt{19}}{12},\qquad n=\frac{2}{3}.
$$
The solutions of the hypergeometric equation are well known. This equation has three regular singular points at $z=0$, $z=1$ and at $z\to \infty$.

       Once the functions $a$ and $b$ are determined, the third equation (\ref{van1}) is not difficult to solve. This equation gives $c_{yy}$ as a function of $y$ and it is straightforward to find $c(y)$ by a double integration. Nevertheless, as it will be seen below, this calculation is irrelevant and only the value of $c_{yy}$ is of importance.

\subsection{Local form of the Calabi-Yau metric}

  After solving the system defining the functions $a(y)$, $b(y)$ and $c(y)$ determining our Calabi-Yau geometries, the next problem is to find the local form of the generic CY metric (\ref{gonoro}) for this case. The Kahler form of the base 4-metric is given by
$$
\omega(y)=\widetilde{\omega}_1-dd^cG=\frac{i}{2} \;(\delta_{i \overline{j}}-4 \;G_{i\overline{j}})\;dz_i\wedge d\overline{z}_j
$$
\be\lb{dv}
=\frac{i}{2}\;(1-4\;G_u-4\;z_i\overline{z}_i \;G_{uu})\;dz_i\wedge d\overline{z}_i-2\;i\;G_{uu}\;(\overline{z}_1\;z_2 \;dz_1\wedge d\overline{z}_2+\overline{z}_2\;z_1 \;dz_2\wedge d\overline{z}_1).
\ee
This expression may be simplified by use of (\ref{postu}) together with the definition $\xi=b-1/4$ to give
\be\lb{dv2}
\omega(y)=-\frac{i}{2}\;(4 \;a\; u+4 \;\xi+4\;a\;z_i\overline{z}_i)\;dz_i\wedge d\overline{z}_i-2\;i\;a\;(\overline{z}_1\;z_2 \;dz_1\wedge d\overline{z}_2+\overline{z}_2\;z_1 \;dz_2\wedge d\overline{z}_1)
\ee
The Kahler 4-metric corresponding to (\ref{dv2}) is simply
\be\lb{rt2}
g_4(y)=-(4 \;a\; u+4 \;\xi+4\;a\;z_i\overline{z}_i)\;dz_i\otimes d\overline{z}_i-4\;a\;(\overline{z}_1\;z_2 \;dz_1\otimes d\overline{z}_2+\overline{z}_2\;z_1 \;dz_2\otimes d\overline{z}_1).
\ee
It is important to remark that when $r\simeq 0$ one has $u\simeq0$ and $z_i\overline{z}_j\simeq0$, and therefore (\ref{rt2}) behaves as
\be\lb{nocon}
g_{4}(y)\simeq -4\;\xi\;(dz_1\otimes d\overline{z}_1+dz_2\otimes d\overline{z}_2).
\ee
Clearly (\ref{nocon}) is, up to an $a$-dependent conformal factor, the flat metric on $R^4$. Therefore near $r=0$ no conical singularity appear. A more handy expression for the (\ref{rt2}) can be obtained by introducing the polar coordinates ($r$, $\theta$, $\psi$, $\phi$) for $R^4$
\be\lb{rv3}
z_1=r \cos\frac{\theta}{2} \;\exp(\frac{i\;\psi+i\;\phi}{2}),\ee
\be\lb{rvr}
 z_2=r \sin\frac{\theta}{2} \;\exp(\frac{i\;\psi-i\;\phi}{2}).
\ee
The expression for the metric (\ref{rt2}) in these coordinates is
\be\lb{familiar}
g_{4}(y)=-\;4(2\;a\;r^2+\xi)\;(\;dr^2+r^2\;\sigma_3^2)-4\;(a\;r^2+\xi)\;r^2\;(\sigma_1^2+\sigma_2^2),
\ee
with $\sigma_i$ being the usual $SO(3)$ Maurer-Cartan forms
$$
\sigma_1=\frac{1}{2}\;(\sin\psi\;d\theta-\sin\theta\;\cos\psi\;d\phi),
$$
\be\lb{morer}
\sigma_2=\frac{1}{2}\;(-\cos\psi\;d\theta-\sin\theta\;\sin\psi\;d\phi),\ee
$$
\sigma_3=\frac{1}{2}\;(d\psi+\cos\theta\;d\phi),
$$
for which the angles, a priori, take values in the following interval
$$
0 \leq \theta \leq\pi,
$$
\be\lb{priori}
0\leq\phi\leq 2\;\pi,
\ee
$$
 0\leq\psi\leq4\;\pi.
$$
Clearly, when $r\simeq 0$ the metric (\ref{familiar}) is approximated by
\be\lb{nocono2}
g_4(y)=-4\;\xi\;\bigg(dr^2+r^2 (\sigma^2_1+\sigma^2_2+\sigma^2_3)\bigg),
\ee
which gives another check that non conical singularity appears near $r=0$. On the other hand, the fiber metric of the general Calabi-Yau metric (\ref{gonoro}) is determined in terms of $G_{yy}$ namely
\be\lb{fc}
G_{yy}=\frac{a_{yy}}{2}\;u^2+b_{yy}\;u+c_{yy}.
\ee
The explicit value of the derivatives appearing in (\ref{fc}) are obtained by use of (\ref{van1}), the result is
\be\lb{ut2}
a_{yy}=64\;a^2,\qquad
b_{yy}=48\;a\;\xi.
\ee
\be\lb{ut3}
c_{yy}=16\;(\xi-\frac{1}{4})\;(\xi+\frac{1}{4})+1=16\;\xi^2.
\ee
Inserting the  expressions (\ref{ut2})-(\ref{ut3}) into (\ref{fc}) gives
\be\lb{landa2}
G_{yy}=32\;(a\;u)^2+48\;\xi\;(a\;u)+16\;\xi^2,
\ee
which can be expressed more neatly as
\be\lb{landa3}
G_{yy}=16\;(2\;a\;u+\xi)(a\;u+\xi).
\ee
Finally, the connection $A$ defining our bundle is given by
\be\lb{con}
A=d^cG_y= G_{yu} d^cu=i\;(a_y\;u+b_y)\;(\overline{z}_1 dz_1-z_1 d\overline{z}_1+\overline{z}_2 dz_2-z_2 d\overline{z}_2),
\ee
where we have used (\ref{lula}) to calculate the action of $d^c$ over ($z_i$, $\overline{z}_i$).  By use of the parameterizations (\ref{rv3}) and (\ref{rvr}) we may reexpress (\ref{con}) as
\be\lb{con2}
A=(a_y\;r^2+\xi_y)\;r^2\;\sigma_3,
\ee
where the relation $b_y=\xi_y$ has been used.  Taking into account the general local form for our Ricci-flat Kahler metrics (\ref{gonoro})  together with the formulas (\ref{verver2}), (\ref{verver3}), (\ref{hoerv}),  (\ref{familiar}),  (\ref{landa3}) and (\ref{con2}) it follows that the local form of the Calabi-Yau metrics we are looking for is given by
$$
g_6=4\;(2\;a\;r^2+\xi)\;(\;dr^2+r^2\;\sigma_3^2)+4\;(a\;r^2+\xi)\;r^2\;(\sigma_1^2+\sigma_2^2)+\frac{16\;(2\;a\;u+\xi)(a\;u+\xi)}{m-\frac{128\;a^3}{3}}\;da^2
$$
\be\lb{pror}
+\frac{1}{16\;(2\;a\;u+\xi)(a\;u+\xi)}\;\bigg(d\alpha+\sqrt{m-\frac{128\;a^3}{3}}(r^2+\xi_u)\;r^2\;\sigma_3\bigg)^2,
\ee
where we made the redefinitions $a\to -a$ and $\xi\to-\xi$. The curvature tensor is irreducible for all the cases described in (\ref{pror}), therefore these metrics are of holonomy exactly SU(3).

\section{Global analysis of the new CY metrics}

    It will be convenient to give a consistency check about the correctness of (\ref{pror}). If we assume intuitively that the radius $r$ has length dimensions $[r]=L$ and that the metric has dimensions $L^2$ , then clearly
the quantity
 $$
a\;r^2+\xi,
 $$
appearing in the component $g_{rr}$, should be dimensionless. This will be the case if $[a]=L^{-2}$. Under this assumption all the components of the Kahler base 4-metric of (\ref{pror}) are of dimensions $[L]^2$. Let us see if this is true for the fiber components. The component $g_{yy}$ is clearly dimensionless and therefore we should have $[y]=L$.  On the other hand, the relation between $y$ and $a$ can be read off from (\ref{uve}) and is
 $$
 [y]=[a]^{-\frac{1}{2}}=L,
 $$
 which confirm what we postulated. Moreover the fiber contains the term
 $$
 (a_y\;r^2+\xi_y)\;r^2,
  $$
and it is seen that
$$
[(a_y\;r^2+\xi_y)\;r^2]=L^2 [y]^{-1}=L.
$$
Therefore if $[\alpha]=L$ the whole metric has dimensions $L^2$, which give a consistency check for (\ref{pror}).

     Let us also note that metrics (\ref{pror}) can never be compact if $K=\partial_{\alpha}$ is a global isometry. This is because a Killing vector satisfy the equation
   $$
   -\Box K_i+R_{ij}K^j=0,
   $$
   and by multiplying this equation by $K_j$ and integrating over the manifold gives that
   $$
   \int_{M_6} (|\nabla_i K_j|^2-R_{ij}K^i K^j).
   $$
   Here the first term is obtained by integration by parts and there is a boundary term which vanish if $M_6$ is compact. If the manifold is Ricci flat then $|\nabla_i K_j|^2=0$ pointwise in the manifold, and such vectors do not exist in a compact manifold.

      In order to make a global analysis of  the metric (\ref{pror}) there are several
      options to consider, depending the signature of (\ref{pror}) and the sign of the parameter $m$ defined in (\ref{uve}). For $m>0$ the function $a$ may take positive or negative values, as shown in (\ref{belo}), with maximal value $a_0$ given $128\;a_0^3=3\;m$. In general the range of $u=r^2$ will be constrained unless $a>0$ and $\xi>0$, in this case the range of $r$ is (0, $\infty$). Clearly there will be a power law singularity in the metric when $(2\;a\;u+\xi)(a\;u+\xi)=0$. In principle one may choose the hypergeometric function $\xi(a)$ in such a way that is always positive and never zero in the range of values of $a$. It seems that in this case this singularity will be absent. The problem with this argument is that the metric (\ref{pror}) does not have an asymptotic region where its is locally flat, so it is should have a singularity, like a delta in the curvature acting as a cosmological constant. Therefore even if we manage to construct such a strictly positive function $\xi(a)$ in our interval this does not mean that the singularity is not present, but that we are using a bad coordinate system. For instance, the metric $dr^2+r^3 d\theta^2$ is singular, but with a parametrization $r=x+1/x$ the singularity would not be seen for any positive $x$ value. Arguably, the same situation is happening here. Therefore none of the constructed metrics should be geodesically complete.

  \section{Discussion}

     To conclude we remark the most salient features of the present work. We have developed a method to construct six dimensional Calabi-Yau metrics in terms of an initial 4-dimensional hyperkahler structure generalizing the situations considered in \cite{Fayyazuddin}, the last ones correspond to D6 branes with two directions inside a complex submanifold of the hyperkahler space and one direction transversal to it parameterized with a coordinate $y$. In all the cases the resulting CY metrics posses a Killing vector which preserve the whole $SU(3)$ structure.

      It should be remarked that in general the CY metrics we described are \emph{not}  fibered over the initial hyperkahler metric. This situation in fact occurs in \cite{Fayyazuddin} and the physicist explanation is that when the D6 branes wrap the complex submanifold their back-reaction deform the initial hyperkahler geometry. The same consideration follows here and the resulting CY metrics are fibered over a 4-dimensional base which is just Kahler and depends on the coordinate $y$ as a parameter. In other words it is Kahler for any fixed value of $y$.

      We also showed the the entire geometry is described in terms of a single function $G$ satisfying a non linear quadratic differential equation (\ref{munja2}). That this function is enough to
 determine the full geometry is evident, as the whole SU(3) structure (\ref{nova3}) is defined purely in terms of $G$ and derivates. The function $G$ depends on $y$ and in fact can be interpreted as the Kahler potential for the base 4-dimensional metric for any fixed value of $y$.  We studied the non linear problem associated to the flat hyperkahler metric in $R^4$ and we found some particular solutions, which allowed us to construct a family of CY examples with power law singularities.

     An interesting open problem is the one arising by fact that  the supergravity solution describing a D6-branes wrapping the $S^2$ in  Eguchi-hanson gravitational instanton is described by the resolved conifold metric \cite{Gomis}-\cite{Edelstein2}. As the Eguchi-Hanson metric is hyperkahler, the method developed here may be suitable to make this link explicit. In particular
it may be one of the situations for which the Fayyazuddin linearization works properly. We hope to find an answer to these questions in the near future.
\\

{\bf Acknowledgement:} The completion of the present work would not be possible without several important advices given by S. Cherkis, who also participated at early stages of this project. Also I sincerely acknowledge H. Lu, who pointed out a crucial numerical error in the initial form of this work. I have also been benefited with short but stimulating discussions with J. Vazquez-Poritz, M. Leston and C. Saemann.  This work is supported by the Science Foundation Ireland Grant N06/RFP/MAT050.
\\

\end{document}